\documentclass[12pt,preprint]{emulateapj}

\newcommand{\ltsima}{$\buildrel<\over\sim$}
\newcommand{\lapprox}{\lower.5ex\hbox{\ltsima}}
\newcommand{\msun}{M$_{\odot}$}
\newcommand{\kband}{{\it K$_s$}-band}

\shorttitle{Quenching of Star Formation and AGNs}
\shortauthors{Bundy et al.}
\slugcomment{submitted to ApJ}

\begin{document}

\title{AEGIS: New Evidence Linking Active Galactic Nuclei to the Quenching of
  Star Formation}

\author{Kevin Bundy\altaffilmark{1}, Antonis
  Georgakakis\altaffilmark{2,3}, Kirpal Nandra\altaffilmark{2}, Richard
  S. Ellis\altaffilmark{4}, Christopher J. Conselice\altaffilmark{4,5},
  Elise Laird\altaffilmark{2}, Alison Coil\altaffilmark{6,7}, Michael
  C. Cooper\altaffilmark{6,8}, Sandra M. Faber\altaffilmark{9}, Jeff
  A. Newman\altaffilmark{10}, Christy M. Pierce\altaffilmark{11}, Joel
  R. Primack\altaffilmark{11}, Renbin Yan\altaffilmark{10}}

\altaffiltext{1}{Reinhardt Fellow, Dept.~of Astronomy and Astrophysics, University of Toronto, 50 St.~George Street, Rm 101,
  Toronto, ON M5S 3H4, Canada}

\altaffiltext{2}{Astrophysics Group, Blackett Laboratory, Imperial College, London SW7 2AZ, UK}

\altaffiltext{3}{Marie Curie Fellow}

\altaffiltext{4}{105--24 Caltech, 1201 E. California Blvd., Pasadena, CA 91125}

\altaffiltext{5}{University of Nottingham, School of Physics \& Astronomy, Nottingham, NG72RDUK}

\altaffiltext{6}{Steward Observatory, University of Arizona, 933 N.\ Cherry Avenue,
 	Tucson, AZ 85721 USA} 

\altaffiltext{7}{Hubble Fellow}
\altaffiltext{8}{Spitzer Fellow}

\altaffiltext{9}{University of California Observatories/Lick Observatory, Board of Studies in Astronomy and Astrophysics,
  University of California, Santa Cruz, CA 95064}
  
\altaffiltext{10}{Department of Astronomy, University of California at Berkeley, MC 3411, Berkeley, CA 94720}

\altaffiltext{11}{Department of Physics, University of California,
    Santa Cruz, 1156 High Street, Santa Cruz, CA 95064, USA}

\begin{abstract}

  Utilizing {\it Chandra} X-ray observations in the All-wavelength Extended
  Groth Strip International Survey (AEGIS) we identify 241 X-ray
  selected Active Galactic Nuclei (AGNs, $L_{2-10{\rm keV}} > 10^{42}$
  ergs s$^{-1}$) and study the properties of their host galaxies in the
  range $0.4 < z < 1.4$.  By making use of infrared photometry from
  Palomar Observatory and $BRI$ imaging from the Canada--France--Hawaii
  Telescope, we estimate AGN host galaxy stellar masses and show that
  both stellar mass and photometric redshift estimates (where necessary)
  are robust to the possible contamination from AGNs in our X-ray
  selected sample.  Accounting for the photometric and X-ray sensitivity
  limits of the survey, we construct the stellar mass function of X-ray
  selected AGN host galaxies and find that their abundance decreases by
  a factor of $\sim$2 since $z \sim 1$, but remains roughly flat as a
  function of stellar mass.  We compare the abundance of AGN hosts to
  the rate of star formation quenching observed in the total galaxy
  population.  If the timescale for X-ray detectable AGN activity is
  roughly 0.5--1 Gyr---as suggested by black hole
  demographics and recent simulations---then we deduce that the inferred AGN ``trigger'' rate
  matches the star formation quenching rate, suggesting a link between
  these phenomena.  However, given the large range of nuclear accretion
  rates we infer for the most massive and red hosts, X-ray selected AGNs
  may not be {\em directly} responsible for quenching star formation.

\end{abstract}


\section{Introduction}\label{intro}

Recent observations of the galaxy population and its evolution since $z
\approx 2$ reveal a pattern in which the most massive
galaxies appear to shut down star formation activity at early times
with increasingly less massive galaxies following later
\citep[e.g.,][]{juneau05, treu05, bundy06,
  borch06, cimatti06}.  This pattern of ``quenching'' in the star
formation history of galaxies---thought to be largely responsible for
the growing abundance of galaxies on the red sequence
\citep[e.g.,][]{faber07}---is commonly referred to as ``downsizing''
\citep{cowie96}.  


Given that the dark matter halos of galaxies are expected to assemble
hierarchically in the $\Lambda$CDM paradigm, understanding the physical
mechanisms responsible for downsizing remains an important challenge.
We seek a process capable of quenching star formation, driving galaxies
onto the red sequence, and preventing further star formation.  Work with
the DEEP2 Galaxy Redshift Survey \citep{davis03} has shown that such a
process must operate over a range of environmental densities
\citep{bundy06, cooper07, gerke07}, suggesting an internal component to
quenching that acts in addition to the suppression of star formation
expected in high-density environments.  Meanwhile, theoretical work has
focused on one such internal process, namely the potential role played
by AGN feedback \citep{silk98}, both as a way of explosively initiating
the quenching event \citep{granato04, scannapieco05, hopkins05b} and
preventing hot gas in already passive galaxies from cooling to form
stars \citep{croton06, bower06, cattaneo06, de-lucia07}.

These scenarios remain largely untested because observational evidence
has been difficult to obtain.  One of the most promising ways forward is
to examine the properties of AGN host galaxies and search for signatures
of this feedback.  Such observations are challenging however because no
selection method finds all galaxies that host active nuclei
\citep{mushotzky04} and brighter AGNs (quasars) can easily outshine
their hosts.  With such difficulties in mind, previous studies suggest
that most lower luminosity AGNs tend to be found in massive, mostly
spheroidal galaxies \citep{dunlop03, kauffmann03b, grogin05, pierce07}
although some ``transition'' sources exhibit disturbed morphologies
\citep[e.g.,][]{canalizo01, hutchings06, conselice07}.  AGNs detected
through emission line diagnostics and X-rays appear to prefer host
galaxies on the red sequence and ``green valley'' \citep{nandra07,
  martin07, salim07}.


Further progress on testing the link between AGN activity and the
downsizing of star formation requires that we understand the stellar
mass distribution and redshift evolution of AGN hosts as compared to
evolutionary patterns in the general galaxy population.  In this paper
we use deep {\em Chandra} observations to identify AGN hosts in the
AEGIS field \citep{davis07} where spectroscopic and photometric
redshifts (photo-$z$'s) as well as infrared photometry provide reliable
stellar mass estimates out to $z = 1.4$.  We then compute the AGN host
stellar mass function and use it to estimate the rate at which AGN
activity is triggered in the galaxy population.  We will show that this
mass-dependent rate is consistent with the rate of star formation
quenching of all galaxies at the same epochs at which the AGN activity
is observed.


The structure of the paper is as follows.  We describe the
multi-wavelength observations and properties of the sample in
$\S$\ref{data} and $\S$\ref{samples}.  The way in which stellar masses
and restframe colors are determined is given in $\S$\ref{methods} while
our methods for constructing mass functions and the resulting AGN host
mass function are presented in $\S$\ref{mfn}.  In
$\S$\ref{discussion} we present evidence for a link between AGN activity
and quenching in the context of estimates of the X-ray AGN timescale and
explore whether feedback from X-ray selected AGNs causes quenching.  We
summarize in $\S$\ref{summary}.  Where necessary, we assume a standard
cosmological model with $\Omega_{\rm M}=0.3$, $\Omega_\Lambda=0.7$,
$H_0=70 h_{70}$ km~s$^{-1}$~Mpc$^{-1}$.

\section{Observations}\label{data}

In this section, we discuss the various observations we use to
investigate the link between AGNs and galaxy evolution.  We begin by
summarizing the DEEP2 Galaxy Redshift Survey \citep{davis03} and a
followup infrared imaging campaign conducted at Palomar Observatory.
One of the four fields in this survey, the Extended Groth Strip (EGS),
was imaged with {\it Chandra} as part of the AEGIS program, providing a
sample of AGN host galaxies that we utilize here.  We will compare the
evolution of this AGN host sample to trends observed in galaxies drawn
from the full DEEP2/Palomar survey.  As discussed in \citet{bundy06},
this sample provides robust stellar mass estimates that can be used to
investigate mass-dependent evolution out to $z = 1.4$.



\subsection{The DEEP2/Palomar Survey}

We provide a brief summary of the DEEP2 Galaxy Redshift survey and the
near-infrared (IR) followup imaging conducted at Palomar Observatory.
The data sets involved were utilized by \citet{bundy06} to explore the
nature of star formation downsizing since $z \approx 1.2$, and further
details are provided there.  Now complete, the DEEP2 Galaxy Redshift
survey \citep{davis03} utilized DEIMOS on the Keck-II telescope to
obtain spectroscopic redshifts for $\sim$40,000 galaxies with $z
\lesssim 1.5$.  The survey is magnitude limited at $R_{AB} \leq 24.1$
and covers more than 3 square degrees over four fields, one of which is
the AEGIS field and is described further below.

Redshift targets were selected using $BRI$ photometry from the
Canada--France--Hawaii Telescope (CFHT) with the 12K$\times$8K mosaic
camera \citep{cuillandre01}.  As described in \citet{coil04}, the
photometry reaches $R_{AB} \sim 25.5$ and was also used for estimating
photo-$z$'s and restframe $(U-B)$ colors, as discussed below.
In the three non-AEGIS fields, target selection in DEEP2 was carried out
using observed colors to exclude sources with $z
< 0.7$.  These selection criteria successfully recover 97\% of the
$R_{AB} \leq 24.1$ population at $z > 0.75$ with only $\sim$10\%
contamination from lower redshift galaxies \citep{davis05}.  More
details on the observing strategy and characteristics of the DEEP2
sample are provided in \citet{coil04}, \citet{willmer06},
\citet{davis05}, \citet{faber07}, \citet{davis07}, and Newman et al.~(in preparation).

Followup imaging of the DEEP2 survey in the $J$ and \kband\ were carried
out using Wide Field Infrared Camera \citep[WIRC,][]{wilson03} on the 5m
Hale Telescope at Palomar Observatory.  These observations are discussed
in detail in \citet{bundy06}.  Excluding the AEGIS field, which was the
highest priority, 0.9 square degrees of DEEP2 were imaged primarily in
the \kband\ with exposure times varying from 2 to 8 hours, depending on
the conditions, and the final typical 80\% completeness depth of
$K_{AB}=21.5$.

We used SExtractor \citep{bertin96} to detect and measure \kband\
sources and cross-referenced them with the CFHT optical and DEEP2
redshift catalogs to construct a K-selected sample that forms the basis
of our analysis.  For total magnitudes used to estimate stellar
masses, we took the {\sc MAG\_AUTO} output from SExtractor and did not
correct this Kron-like magnitude for missing light.  We estimated the
uncertainty on these magnitudes by inserting fake sources at various
magnitudes and using SExtractor to recover them.  Based on the locations
of K-selected sources, we measured aperture photometry in the $BRI$ data
using 2\arcsec\ diameter apertures which we found exhibited the least
scatter.  These colors are used in fitting galaxy spectral energy
distributions (SEDs) needed for both photo-$z$'s and stellar
mass estimates.

\subsection{The AEGIS Field}

Covering the 0.5 deg$^2$ Extended Groth Strip ($\alpha = 14^{\rm
  h}17^{\rm m}, \delta = +52^{\circ}30$\arcmin), the All-Wavelength
Extended Groth Strip International Survey \citep[AEGIS,][]{davis07}
accounts for one of the four fields where DEEP2 redshifts and Palomar
near-IR imaging were obtained, although there are slight differences
with respect to the other DEEP2 fields (above) that we describe here.
This field is also special because many other observatories, including
the {\it Hubble} and {\it Spitzer Space Telescopes}, have conducted
observations there.  Of key importance for this work are the deep X-ray
data from the Advanced CCD Imaging Spectrometer (ACIS) on {\it Chandra}.

The {\it Chandra} data are described in \citet{georgakakis06} and
\citet{georgakakis07}, and the X-ray data analysis is presented in
\citet{nandra05}.  Briefly, the AEGIS region was targeted over several
epochs for a total integration time of 190 ks.  Standard reduction
methods using the {\sc CIAO} software were employed to derive fluxes in
four energy bands---0.5--7.0 keV (full), 0.5--2.0 keV (soft), 2.0--7.0
keV (hard), and 4.0--7.0 (ultrahard)---by integrating the counts detected
within the 70\% encircled energy radius at each source position.  The
counts in each observed band were converted to the standard bands of
0.5--10, 0.5--2, 2--10, and 5-10 keV by assuming an intrinsic power-law
with $\Gamma = 1.4$ and Galactic absorption.  The typical detection
limits in each band are 35, 1.1, 8.2, and 14 in units of $10^{-16}$ ergs
s$^{-1}$ cm$^{-2}$, although these depend on position because the {\it
  Chandra} sensitivity declines away from the center of each pointing.
X-ray detections were referenced to the optical/IR photometric catalogs
using the method described in \citet{georgakakis06}.

In terms of the DEEP2 spectroscopic redshifts, target selection was
carried out differently in AEGIS than it was in the other DEEP2 fields.
Here, a straight magnitude limit of $R_{AB} \leq 24.1$ was employed with
galaxies having $z \lesssim 0.7$ down-weighted but not excluded from the
target sample.  The near-IR Palomar imaging is deepest in EGS,
especially along the center of the strip where $J$-band imaging was also
obtained, and the typical 80\% completeness depth is $K_{AB}=22.5$.
Overall, the typical AEGIS depth reaches $K_{AB} \approx 22$.  The
properties of the CFHT $BRI$ data are identical to those described
above.

\section{The Galaxy and AGN Host Samples}\label{samples}

To begin our analysis of AGN hosts and their relationship to evolution
in the galaxy population we distinguish between the {\it galaxy
  sample} and the {\it AGN host sample}.  The galaxy sample is comprised of the full DEEP2/Palomar survey, including the
AEGIS.  With the goal of providing a self-consistent benchmark for AGN
host comparisons, we note that this sample is slightly larger than the
one presented in \citet{bundy06} because unlike that paper, our analysis
does not require that we restrict the perimeter of the survey on account
of accurate environmental density measurements.  While X-ray
observations are not required for the galaxy sample,  we do select sources
in the \kband\ with good quality \citep[``zquality'' $\geq 3$, see][]{davis07} spectroscopic
redshifts in the redshift range $0.4 < z < 1.4$, reaching \kband\ limits that
depend on redshift.  As in \citet{bundy06}, we use $K_{AB} \leq
21.8, 22.0, 22.2$ for $z \leq 0.7, 1.0, 1.4$.  


We apply the same limits to the AGN host galaxies which are identified
in the {\it Chandra} full band and have a Poisson false detection
probability that is less than $4 \times 10^{-6}$.  We then use the redshift
of the matched host and assume $\Gamma = 1.9$ to
infer $L_{2-10}$, the X-ray luminosity in the 2--10 keV energy band.
This effectively corrects for absorption for column densities of $N_H
\lesssim 10^{23}$ cm$^{-2}$ at $z \approx 1$ \citep{nandra94}, and we apply this
procedure even if the source is not detected in the hard band.  

AGNs are identified as those sources with $L_{2-10} > 10^{42}$ ergs
s$^{-1}$; we note that AGNs are certainly present in sources below this
threshold, although they are more difficult to distinguish with the present
data.  Because normal galaxies without AGNs have not been observed with
$L_{2-10} > 10^{42}$ ergs s$^{-1}$ \citep[see][]{bauer04}, we adopt this limit
to ensure that there is little to no contamination from sources without AGNs.
Indeed, even without an X-ray luminosity threshold, the flux limit of the
Chandra data ($8.2 \times 10^{-16}$ ergs s$^{-1}$ cm$^{-2}$ in the hard band)
suggests a very low non-AGN contamination rate of only a few percent in the
observed source densities \citep{bauer04}.  

While hard X-ray selection provides one of the cleanest ways to select
AGNs at high redshift \citep[e.g.,][]{mushotzky04, barger05}, it will
still miss Compton-thick sources ($N_H \gtrsim 10^{24}$ cm$^{-2}$),
which, while not expected to outnumber hard X-ray detected AGNs, are
thought to contribute at least half as many \citep[e.g.,][]{treister06,
  guainazzi05, gilli07}.  We will account for missed Compton-thick AGNs
by incorporating the detection efficiency of our sample in the analysis
in $\S$\ref{discussion}.

In what follows, we will consider the AGN host sample with spectroscopic redshifts
and $R_{AB} \leq 24.1$ (the spec-$z$ sample, 84 sources) as well as a deeper sample
($R_{AB} \leq 25.1$) of AGN hosts supplemented with photometric
redshifts (the photo-$z$ sample, 241 sources).  As in \citet{bundy06}, photometric
redshifts in AEGIS were estimated using two methods.  For galaxies with
$R_{AB} \leq 24.1$ we use ANN$z$ \citep{collister04}, a neural network
redshift estimator, which benefits from a large training set provided by
the magnitude-limited nature of the DEIMOS target selection in the EGS.
We also include host galaxies with $24.1 < R_{AB} \leq 25.1$ and
3$\sigma$ detections in the $BIK$ bands, but use the BPZ SED-based
estimator \citep{benitez00} for these sources.  By comparing these
photo-$z$'s to the DEEP2 spectroscopic redshifts we find $\delta z
/(1+z) = 0.11$.  Further details on these two methods as applied to
galaxies in the DEEP2/Palomar sample can be found in \citet{bundy06}.
Photometric redshifts in the EGS based on the CFHT Legacy Survey are
also available from \citet{ilbert06}.  Comparing these estimates to the
DEEP2 spectroscopic redshifts we find $\delta z /(1+z) = 0.17$ and
therefore use the slightly better ANN$z$+BPZ estimates described above.  Note that
\cite{ilbert06} redefine $\delta z /(1+z)$ as $1.48 \times {\rm median}
\left| z_{\rm spec} - z_{\rm phot} \right| / (1 + z_{\rm spec})$ and that with
this definition the comparison to DEEP2 yields $dz_{\rm Ilbert} = 0.03$,
similar to the accuracy reported in \citet{ilbert06}.

\begin{figure}
\plotone{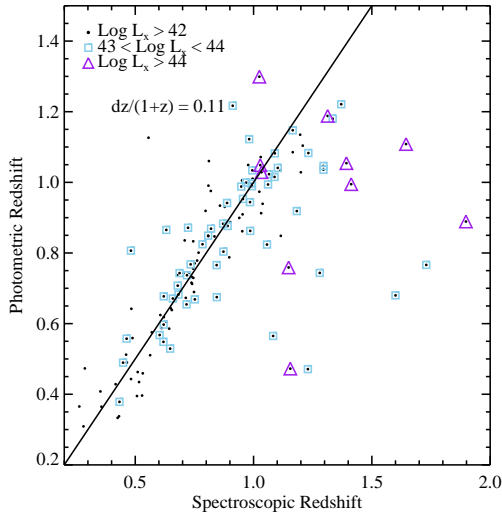}
\caption{Comparison of photo-$z$ quality for the AGN spec-$z$ host
  sample.  Four additional outliers with $z_{\rm spec} > 2.0$ are not
  shown, and all have $L_{2-10} > 10^{43}$ ergs s$^{-1}$, with three having $L_{2-10} >
  10^{44}$.  As indicated, for $0.4 < z_{\rm phot} < 1.4$, $\delta z
  /(1+z) = 0.11$, the same as the full galaxy population.  We note, however, that
  photo-$z$ outliers tend to be the most X-ray luminous.
  \label{fig:zphot_comp}}
\end{figure}

An obvious concern with photometric redshift estimates for AGN hosts is
that non-thermal contamination  could lead to large
redshift errors.  By comparing photometric and spectroscopic redshift
estimates in the AGN spec-$z$ host sample in Figure \ref{fig:zphot_comp}
we show that in general this is not the case.  Since the global sample of galaxies ($0.4 < z_{\rm phot} < 1.4$) is characterized by the
same $\delta z /(1+z) = 0.11$, this comparison demonstrates that
photometric redshifts of X-ray selected AGN hosts can be believed at the
same level as those of non-AGN galaxies.  This implies that AGN contamination in
the optical/IR is not significant, as expected \citep[e.g.,][]{barger05}.  Figure
\ref{fig:zphot_comp} does show, however, that the photo-$z$ outliers are
dominated by the more X-ray luminous systems, suggesting more
contamination in these cases.  Caution must be used in interpreting
the properties of these sources.

\begin{figure}
\plotone{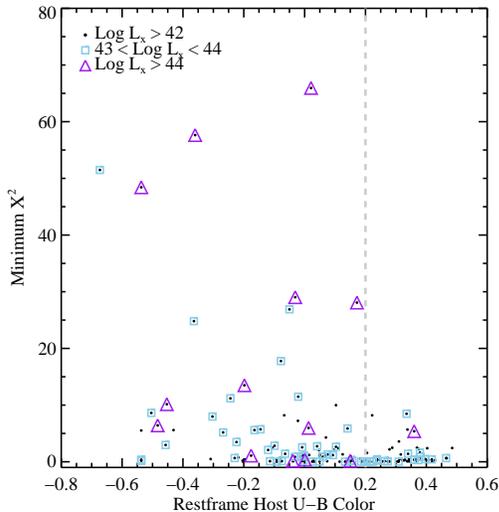}
\caption{Reduced $\chi^2$ of the best fitting SED determined by the
  $M_*$ estimator as a
  function of restframe ($U-B$) color.  More X-ray luminous AGNs tend to
  exhibit poorer fits, indicating the presence of AGN contamination.
  The blue colors of the AGN hosts of X-ray luminous sources may be
  caused in part by this contamination.  The grey dashed line is typical
  of the division used to divide red and blue galaxies.
  \label{fig:chi_mass}}
\end{figure}

\section{Physical Properties}\label{methods}

\subsection{Restframe $U-B$ Color}

Following \citet{bundy06} and \citet{nandra07}, we use the methods
described in \citet{willmer06} to estimate the restframe ($U-B$) colors
in both our galaxy and AGN host samples.  This measurement is frequently
used as a diagnostic of star formation activity in galaxies and exhibits
a bimodal distribution to at least $z \sim 1$ \citep{bell04}, separating
star-forming ``blue cloud'' galaxies from the mostly passive ``red
sequence.''  We used the cut employed by \citet{van-dokkum00} which can
be expressed in Vega magnitudes as,

\begin{equation}\label{col_lim}
U - B = -0.032(M_B + 21.52) + 0.454 -0.25
\end{equation}

\noindent The ($U-B$) color distribution for the AGN spec-$z$ host sample
is illustrated in Figure \ref{fig:chi_mass} and was originally discussed
for a subset of the data utilized here in \citet{nandra07}.  The
properties of the current and somewhat larger sample here show
agreement with those presented in \citet{nandra07}.

\subsection{Stellar Mass Estimates}

To estimate stellar masses ($M_*$), we use the methods described in
\citet{bundy06}.  Based on the observed $BRIK$ colors (measured using
2\farcs0 diameter apertures) and the redshift information for each
galaxy, we fit the observed SED to a grid of 13440 models constructed
using the \citet{bruzual03} population synthesis code.  The grid spans a
range of metallicities, star formation histories (parametrized as
exponential), ages, and dust content.  The grid is restricted such that
only models with ages (roughly) less than the cosmic age at a galaxy's
redshift are considered.  Systematic uncertainties are introduced by our
choice of a Chabrier IMF \citep{chabrier03} and the use of the
\citet{bruzual03} code.  While some studies support $M_*$ estimates
based on the \citet{bruzual03} software \citep[see][]{kannappan07},
others have pointed out potential problems \citep{maraston06}.  In
\citet{conselice07}, we show that a preliminary analysis with the latest
Bruzual and Charlot models (2007, in prep.) yield mass estimates that
are $\sim$0.07 dex smaller on average.  This issue will be discussed
further in future work.

With the grid defined in this way, the SED of a given galaxy is then
compared to the model at each grid point, where the specific
\kband~$M_*/L_K$ ratios and $M_*$ values for each model are also stored.
The probability that each model fits the data is then summed or
marginalized over the grid to yield the stellar mass probability
distribution.  The median of this distribution is taken as the estimate
of $M_*$, and the width provides an estimate of the uncertainty,
typically 0.1--0.2 dex.  This is added in quadrature to the \kband\
magnitude uncertainty to determine the final error on $M_*$.  Stellar
mass estimates for galaxies with only photometric redshifts suffer from
the uncertainty in luminosity distance introduced by the photo-$z$
uncertainties and the possibility of catastrophically wrong redshift
information.  A study of the effect of photometric redshifts on stellar
mass was performed in \citet{bundy05}.

For the case of the AGN host samples, it is important to consider the
effects of non-thermal contamination which may affect the inferred
stellar masses.  Including an AGN component in the model SEDs used to
estimate the stellar mass is not practical.  Instead we investigate the
potential error by plotting the reduced $\chi^2$ values of the best
fitting SED from the stellar mass grid for each member of the AGN
spec-$z$ host sample in Figure \ref{fig:chi_mass}.  Even under the best
conditions, we do not require a perfect fit to the observed SED because
our Bayesian mass estimator considers a range of models when assigning
the final mass estimate.  Still, mass estimates with $\chi^2 > 30$
should be considered with caution.  Fortunately, while the average
$\chi^2$ value of AGN hosts is higher than for galaxies without AGNs (e.g.,
the fraction of AGN hosts with $\chi^2 > 10$ is 10\%, while this number
for all galaxies is 1\%) and seems to correlate with $L_{2-10}$, only 4
hosts have $\chi^2 > 30$, indicating that the mass estimates for the AGN
host sample are robust.


\section{Stellar Mass Functions}\label{mfn}

\subsection{Methods}

With the $M_*$ estimates described above, we are now in a position to
construct stellar mass functions (MFs) from our sample.  We will
describe our formalism in this section and present the AGN host mass
function in $\S$\ref{results}.  The same methods are also applied to
construct MFs for the full galaxy sample.  These allow us to study the
growing fraction of red galaxies which we will use to infer the star
formation quenching rate in $\S$\ref{edot}.

Constructing galaxy stellar MFs requires understanding and correcting
for the limitations and completeness of the survey data.  We adopt the
$V_{max}$ formalism \citep{schmidt68} to this end, following
\citet{bundy06}.  In the case of the galaxy sample, the maximum volume
can be limited by either the \kband\ depth or the $R_{AB} \leq 24.1$
limit used to define the DEEP2 spectroscopic sample.  In this case, for
each galaxy $i$ in the redshift interval $j$, the value of $V^i_{max}$
is given by the minimum redshift at which the galaxy would leave the
sample,

\begin{equation}
V^i_{max} = \int_{z_{low}}^{z_{high}} d \Omega_j \frac {dV}{dz} dz
\end{equation}

\noindent where $d \Omega_j$ is the solid angle subtended by the sample
defined by the limiting \kband~magnitude, $K^j_{lim}$ (which changes
depending on the redshift interval $j$), and $dV/dz$ is the comoving
volume element.  The redshift limits are given as,

\begin{equation}
z_{high} = {\rm min}(z^j_{max}, z^j_{K_{lim}}, z_{R_{lim}})
\end{equation}
\begin{equation}
z_{low} = z^j_{min}
\end{equation}

\noindent where the redshift interval, $j$, is defined by $[z^j_{min},
z^j_{max}]$, $z^j_{K_{lim}}$ refers to the redshift at which the galaxy
would still be detected below the \kband\ limit for that particular
redshift interval, and $z_{R_{lim}}$ is the redshift at which the galaxy
would no longer satisfy the $R$-band limit of $R_{AB} \leq 24.1$.  We
use the best-fit SED template as determined by the stellar mass
estimator to calculate $z^j_{K_{lim}}$ and $z_{R_{lim}}$, thereby
accounting for the $k$-corrections necessary to compute accurate
$V_{max}$ values (no evolutionary correction is applied).

In the case of the AGN host samples, the procedure must be modified to
account for the limiting X-ray depth.  This is more complicated because
the limit varies smoothly as a function of position within a given {\it
  Chandra} pointing.  Deeper sensitivity limits correspond to smaller
effective areas.  We therefore compute {\it Chandra} sensitivity
curves in the full band (corresponding to the selection band),
accounting for the overlap with the Palomar near-IR data which is not
complete over the EGS field at all depths.  For a given source with
X-ray luminosity $L_X^i$, we use a $\Gamma = 1.9$ power-law to estimate the observed flux this
source would have as a function of redshift.  We then use the
sensitivity curve to compute the corresponding solid angle over which
such a source could be detected as a function of redshift, $d
\Omega(z)$.  Thus, we derive a second $V_{max}$ estimate for AGN hosts
based on the X-ray limits:

\begin{equation}
V^i_{max,{\rm X-ray}} = \int_{z_{low}}^{z_{high}} d \Omega(z) \frac {dV}{dz} dz
\end{equation}

\noindent where $z_{low}$ and $z_{high}$ are given by the boundaries of
the redshift interval.  The final $V_{max}$ for the AGN hosts is taken
as the smaller of the $V_{max}$ computed based on the $R$ and \kband\
limits and $V_{max, {\rm X-ray}}$.  Typically the X-ray volume provides
the limiting $V_{max}$.

The galaxy sample and AGN spec-$z$ host sample make use of DEEP2
spectroscopic redshifts only.  For this reason, additional weights must
be applied to account for the redshift targeting selection function and
success rate of these samples.  Here we follow the technique described
by \citet{willmer06} and modified in \citet{bundy06}.  Specifically, we
compare the number of sources with good quality redshifts (zquality
$\geq 3$) in a given bin of $(B-R)$/$(R-I)$/$R_{AB}$/\kband\ parameter
space to the total number of sources targeted in that same bin.  We
adopt the ``optimal'' model of \citet{willmer06}, which accounts for the
different ways that red and blue galaxies are likely to be excluded from
the spectroscopic sample.

The situation is more complicated for AGN hosts because applying this
weighting scheme to AGN hosts assumes that the photometric sources in the
corresponding color/magnitude bins also host AGNs.  We therefore modify
the weighting scheme when it is applied to AGN hosts as follows.  In
each redshift interval we determine the ratio between the number of
spectroscopic AGN hosts and the number of potential hosts.  Potential
hosts include galaxies without X-ray detections that have spectroscopic
redshifts, stellar masses greater than the completeness limit, and $U-B
> -0.1$.  This color requirement is motivated by Figure
\ref{fig:chi_mass} which demonstrates that most AGN hosts have such
colors.  For the three redshift intervals $0.4 < z < 0.7$, $0.75 < z <
1.0$, and $1.0 < z < 1.4$, we find AGN fractions of 0.08, 0.09, and 0.14.
The AGN spec-$z$ host weights are then determined by multiplying the
``optimal'' weights discussed above by these numbers and ensuring the
weight does not drop below 1.0.  No weighting is required for the AGN
photo-$z$ host samples; comparisons between the spec-$z$ and photo-$z$
AGN host samples thus provide a useful measure of the success of our
weighting scheme.

\subsection{The AGN Host Stellar Mass Function}\label{results}

\begin{figure*}
\epsscale{0.6}
\plotone{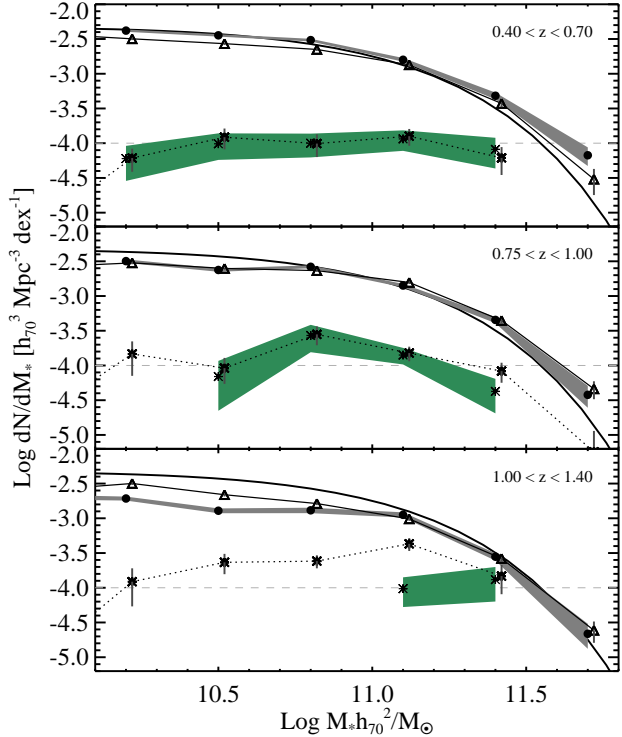}
\caption{The stellar mass function of AGN host galaxies in three
  redshift intervals as compared to the total galaxy stellar mass
  function.  Green shading traces the MF and uncertainty of AGN hosts
  with spectroscopic redshifts and $L_{2-10} > 10^{42}$ ergs s$^{-1}$.
  The asterisk symbols with error bars and connected by the dotted line
  show the AGN host MF for the photo-$z$ supplemented sample.  Total
  mass functions from the AEGIS field are shown with solid circles and
  grey shading (spec-$z$ sample) and triangles (photo-$z$ supplemented
  sample).  The solid line is taken from the best fit of \citet{bundy06}
  to the total MF at $z \approx 0.5$. A horizontal dashed line has been drawn at
  $\phi = 10^{-4}$ in all panels to guide the eye. \label{fig:mfn_x}}
\end{figure*}

We plot the AGN spec-$z$ and photo-$z$ supplemented host mass functions
in Figure \ref{fig:mfn_x} in three redshift intervals.  The green
shading represents the uncertainty in the spec-$z$ sample, arising
primarily from number statistics.  The corresponding mass functions of
the AGN photo-$z$ host sample are indicated by asterisk symbols
connected by dotted lines.  For reference, each panel also indicates the
total spec-$z$ AEGIS mass function (light gray shading with solid
circles), the total photo-$z$ supplemented MF (triangle symbols), and in
all panels the best fitting MF at $z \approx 0.5$ from \citet{bundy06}.
The number of AGN spec-$z$ hosts diminishes significantly in our highest
redshift bin.  Interpretations at these redshifts rely on the photo-$z$
sample only.  Note that the total photo-$z$ sample covers a larger area
(by $\sim$25\%) than the spec-$z$ sample.  Thus, slight differences in
the total mass functions can arise from cosmic variance, especially in
the lowest redshift interval.

\begin{figure}
\plotone{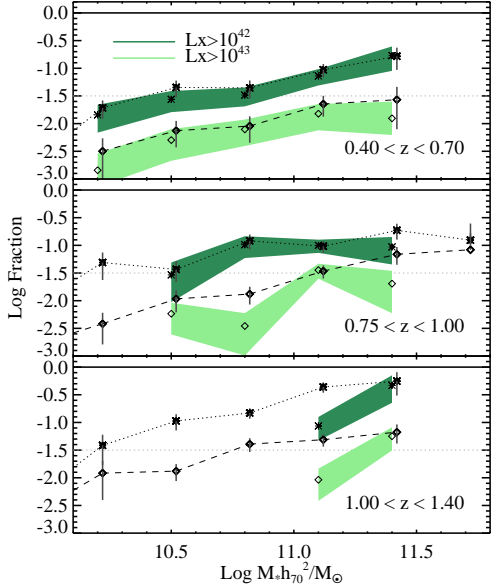}
\caption{Fractional contribution of AGN hosts to the total MF in log units,
  shown in three redshift intervals.  As in Figure \ref{fig:mfn_x}, dark green
  shading denotes the AGN spec-$z$ host sample with $L_{2-10} > 10^{42}$ ergs
  s$^{-1}$, while asterisk symbols connected by dotted lines denote the AGN
  photo-$z$ host sample with the same $L_{2-10}$ threshold.  Light green shading
  corresponds to AGN spec-$z$ hosts with the brighter X-ray cut of $L_{2-10} >
  10^{43}$ ergs s$^{-1}$, while diamond symbols connected by dashed lines show
  the corresponding AGN photo-$z$ MFs.  A dotted horizontal line at
  $\log f = -1.5$ has been drawn in each panel to guide the eye.\label{fig:mfn_xfrac_Lx}}
\end{figure}

At all redshifts, Figure \ref{fig:mfn_x} shows that the AGN host MF is
roughly flat across the stellar mass range sampled.  Comparing the AGN
photo-$z$ MFs, there is evidence that from $z \approx 1.2$ to $z \approx
0.5$ the abundance of AGN hosts decreases roughly by a factor of 2 at
all stellar masses probed.  Because the corresponding number density of
all galaxies is lower at $z \gtrsim 1$, the fraction of systems hosting
AGNs increases at these epochs.  This is shown explicitly in Figure
\ref{fig:mfn_xfrac_Lx} which plots the AGN fraction as a function of
stellar mass.  On one hand, the relatively flat MFs in Figure
\ref{fig:mfn_x} suggest that AGN evolution---for example the declining
hard X-ray luminosity density \citep[e.g.,][]{barger05, hasinger05}---is
independent of host $M_*$.  However, the {\em fraction} of AGN hosts
shown in Figure \ref{fig:mfn_xfrac_Lx} presents a different
interpretation.  As galaxies continue assembling and their abundance grows with time,
X-ray AGNs may be increasingly turning off---especially at the highest masses.
This would lead to stronger evolution in the AGN fraction (Figure
\ref{fig:mfn_xfrac_Lx}) accompanied by milder evolution in the absolute
numbers of AGN hosts (Figure \ref{fig:mfn_x}).

Figure \ref{fig:mfn_xfrac_Lx} also shows the result of splitting the sample
using two different X-ray luminosity thresholds.  The dark green shading and
asterisk symbols denote the AGN host sample with $L_{2-10} > 10^{42}$ ergs
s$^{-1}$ as in Figure \ref{fig:mfn_x}.  The light green shading and diamonds in
Figure \ref{fig:mfn_xfrac_Lx} provide a comparison to the host MF corresponding
to higher X-ray luminosities of $L_{2-10} > 10^{43}$ ergs s$^{-1}$.  Note that
because of the steep decline in the X-ray luminosity function
\citep[e.g.,][]{barger05}, the AEGIS survey area is too small to effectively
sample sources with $L_{2-10} \gtrsim 10^{44}$ ergs s$^{-1}$.

While it appears that the more X-ray luminous AGNs are generally less abundant, we
find little significant difference in the shape of the host MF as a function of
X-ray luminosity.  Those with $L_{2-10} > 10^{43}$ ergs s$^{-1}$ account
for roughly one-third of the full AGN sample with $L_{2-10} > 10^{42}$
ergs s$^{-1}$, but are associated with host galaxies with a similar mass
distribution.  

Studies of the AGN X-ray luminosity function (LF) show that the more luminous
sources are more abundant in the past relative to the less luminous ones, a
phenomenon often termed ``AGN downsizing'' \citep[e.g.,][]{hasinger05,
  barger05}.  Because this trend is most apparent for systems brighter than the
knee in the X-ray LF---that is AGNs with $L_{2-10} \gtrsim 10^{44}$ ergs
s$^{-1}$---we would not expect the effect to be strong in this survey, which is
too small to accurately sample such luminous AGNs.  There is some suggestion,
however, for this effect in the highest mass bin of the $0.7 < z < 1.0$ redshift
interval in Figure \ref{fig:mfn_xfrac_Lx}, where it appears the most massive
hosts become dominated by the brightest X-ray AGNs.


\section{Linking AGNs and Quenching}\label{discussion}

We now move to the primary goal of this paper.  Our aim is to evaluate
the role of AGNs in the evolving star formation properties of the full
galaxy population.  To accomplish this, we will compare the rate at
which AGNs are triggered in galaxies of a given mass with the rate at
which star formation is quenched at these masses.  We will use the AGN
host mass function presented in the previous section (coupled with the
AGN lifetime) to infer the AGN trigger rate.  First, however, we must
characterize the mass-dependent quenching rate in the total population.
We use the full Palomar/DEEP2 sample to provide this measurement below.

\subsection{The Star Formation Quenching Rate}\label{edot}

\begin{figure}
\epsscale{1.2}
\plotone{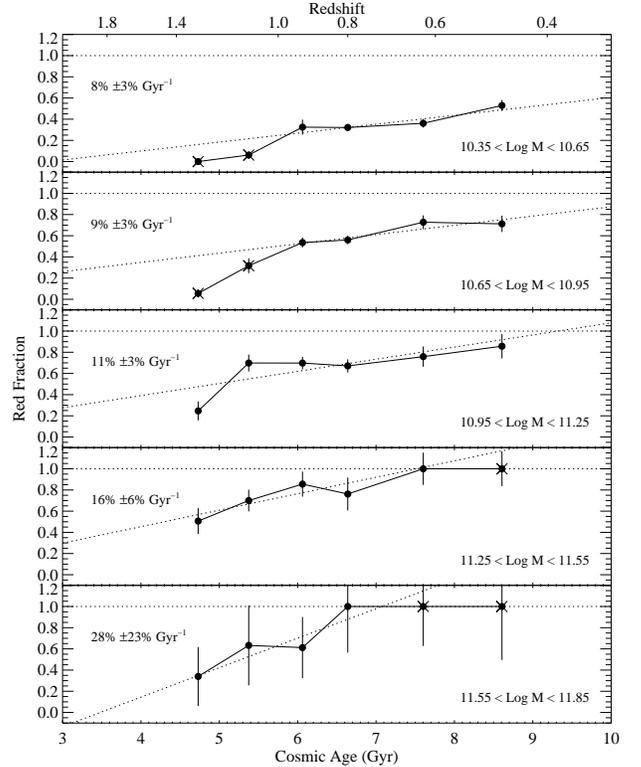}
\caption{Evolving fraction of red galaxies as a function of time in
  various mass bins.  The growing fractions have been fit by the dotted
  lines, excluding data points where the fraction is either 1.0 or the
  data are incomplete.  Excluded data points are indicated by the cross
  symbols.  \label{fig:edot_frac}}
\end{figure}

We will define the quenching rate, $\dot{Q}$, as the fraction of all
galaxies in a stellar mass bin that shift to the red sequence per Gyr.
This quantity can be derived using the methods of $\S$\ref{mfn} to plot
the increasing fraction of red galaxies (relative to the abundance of
all galaxies) as a function of time for various bins of stellar mass
(Figure \ref{fig:edot_frac}).  We will use the slope of the increasing
red fraction to estimate $\dot{Q}$.  We have chosen to study the red galaxy
fraction---as opposed to absolute number densities---because this helps
mitigate uncertainties caused by cosmic variance, which to first order
affect the total number density measured in a given redshift interval
\citep[see][]{bundy06}.

The use of fractional abundances also provides a better handle on the
rate at which galaxies become red, that is $\dot{Q}$.  In the absence of
processes that shift galaxies into different mass bins, quenching only
alters the fraction of red galaxies in a given $M_*$ bin.  The transfer
of galaxies across mass bins is constrained to be small by the lack of
significant evolution in the shape of the total MF from $z \sim 1$.
However, it is possible that both star formation and merging may move
galaxies between mass bins.  As for merging, we make
the assumption that the effect on the red fraction is small if merging
is independent of galaxy type and the merging rate does not vary across
the 0.3 dex mass bins used here.  As for star formation, because lower
mass galaxies exhibit higher SF rates, their evolution would tend to
drive the red fraction down as low mass blue galaxies enter a given mass
bin.  However, as the number of galaxies forming stars at a rate
sufficient to double their mass over a few Gyr is small for $M_* \gtrsim
10^{10}$\msun\ \citep[e.g.,][]{feulner05a}, such an effect would have a
small impact on the red fraction.  Still, we emphasize that the evolving
red fraction in specific mass bins is only an estimate of the true
quenching rate.

The buildup in the fraction of red systems is clear in Figure
\ref{fig:edot_frac} and allows us to crudely fit lines to the fractional
growth rate.  In this fit, we exclude points at early times and low
masses that fall below our expected completeness limit as well as those
points for which the red fraction is equal to 1.0.  We take the slope of these
lines as our estimate of $\dot{Q}$.  In the mass bins centered at $\log
M_*/M_{\odot} = $10.5, 10.8, 11.1, 11.4, 11.7 we find fractional
quenching rates, $\dot{Q}(M_*)$, of 8\%$\pm3$\%, 9\%$\pm3$\%,
11\%$\pm3$\%, 16\%$\pm6$\%, and 28\%$\pm23$\% per Gyr.  While Figure
\ref{fig:edot_frac} shows that our linear approximation adequately fits
the data, we cannot further constrain the quenching rates as a function
of time (or redshift).  Our results are obviously only valid until the
red fraction reaches 1.0 and all systems are quenched and, extrapolating
the fits, we find that ``total quenching'' in these mass intervals
occurs when $z = $ -0.4, 0.0, 0.4, 0.6, 0.7, or in the assumed
cosmology, when the cosmic age equals $\tau_{age} =$ 14.7, 11.4, 9.3,
7.5, 7.3 Gyr.  Figure \ref{fig:edot_frac} not only reinforces the notion
that more massive galaxies become quenched first, but demonstrates the
new result that they may also become quenched {\em faster} than their
lower mass counterparts.

\subsection{Comparing AGN Triggering and Star Formation Quenching}

With the star formation quenching rate measured above, we now utilize
the AGN host mass function to derive the AGN ``trigger rate'',
$\dot{\chi}(M_*)$, defined as the fraction of all
galaxies per Gyr in which X-ray detected AGNs turn on as a function of stellar
mass.  The rate of AGN triggering multiplied by the timescale over which
AGNs are visible at X-ray wavelengths is equal to the observed AGN
fraction (shown in Figure \ref{fig:mfn_xfrac_Lx}).  If we account for
the AGN detection efficiency, $\epsilon$, of our X-ray observations,
we can write this relation as $f_{\rm AGN}(M_*) = \epsilon
\dot{\chi}(M_*) \tau_{\rm AGN}$, where $\tau_{\rm AGN}$ is the X-ray AGN
timescale.

This timescale is the largest uncertainty in the calculation and must be
derived from theoretical arguments.  We will consider three estimates
taken from the literature for the average value of $\tau_{\rm AGN}$.
From the detailed simulations studied in \citet{hopkins05a} we find
$\tau_{\rm AGN} \approx 0.6$ Gyr.  From the statistical and population
arguments for low efficiency AGNs in \citet{marconi04} we use $\tau_{\rm
  AGN} \approx 0.9$ Gyr, and from the model discussed in
\citet{granato04} we use $\tau_{\rm AGN} \approx 1.8$ Gyr.  We will
return to the timescale problem and discuss these estimates further
below.

\begin{figure}
\epsscale{1.2}
\plotone{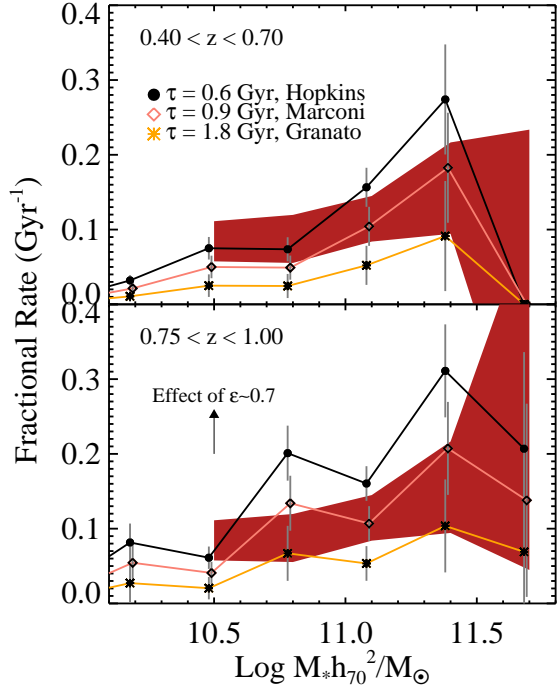}
\caption{Comparison in two redshift bins between the fractional AGN
  ``trigger rates''--- calculated using three estimates of the X-ray AGN
  timescale---and the star formation quenching rate denoted by the red
  shaded region.  The width of the shading illustrates the 1$\sigma$
  uncertainty.  Trigger rates are derived based on the AGN photo-$z$
  sample (similar results are obtained for the spec-$z$ sample),
  assuming all AGNs are detected and X-ray AGN timescales of 0.6, 0.9,
  and 1.8 Gyr based on estimates from the work of \citet{hopkins05a},
  \citet{marconi04}, and \citet{granato04}.  The systematic effect of a
  70\% detection efficiency ($\epsilon \approx 0.7$) is shown by the
  arrow \label{fig:xfrac_tau}}
\end{figure}

Assuming $\epsilon = 1$ we solve for the corresponding trigger rates,
$\dot{\chi}(M_*)$, by dividing the AGN host fraction ($L_{2-10} >
10^{42}$, Figure \ref{fig:mfn_xfrac_Lx}) by the timescales above.  We
plot the results for the two redshift bins where our sample is most
complete in Figure \ref{fig:xfrac_tau}.  The corresponding star
formation quenching rates from our analysis of red galaxies is denoted
by the red shaded region.  The effect of a lower detection efficiency of
$\epsilon \approx 0.7$ (as might arise from missed Compton-thick
sources) increases the trigger rates as roughly shown by the arrow.

Despite the uncertainties and assumptions, Figure \ref{fig:xfrac_tau} demonstrates
surprising agreement in both the normalization and mass dependence of
the rates of quenching and AGN triggering, given the three estimates for
$\tau_{\rm AGN}$.  We interpret this as strong but circumstantial
evidence that the quenching of star formation and AGN activity are
physically related.  We will turn to the question of whether AGNs
actually {\em cause} quenching in $\S$\ref{causality}.

\subsection{The X-ray AGN Timescale}

\begin{figure}
\plotone{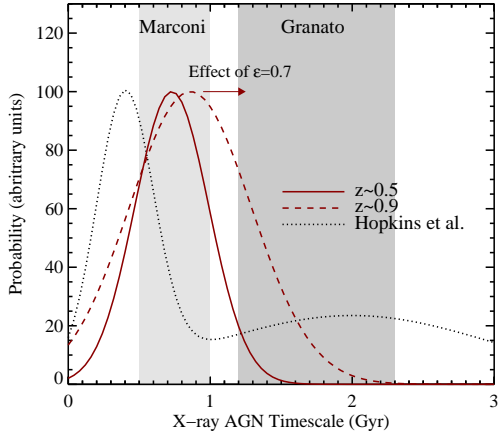}
\caption{Range of X-ray detectable AGN timescales.  As a reference
  point, the result of {\em assuming} the AGN trigger rate equals the
  quenching rate yields timescales in the range indicated by the red
  solid ($z \sim 0.5$) and dashed ($z \sim 0.9$) distributions.  The
  effect of a 70\% AGN detection efficiency ($\epsilon = 0.7$) on these
  distributions is shown by the red arrow.  Independent predictions of
  $\tau_{\rm AGN}$ based on the models of \citet{hopkins05a} have a
  range indicated by the dotted line.  Predictions from
  \citet{granato04} and \citet{marconi04} are denoted by the grey shaded
  regions.  \label{fig:tau}}
\end{figure}

Because the calculation of the AGN trigger rates shown in Figure
\ref{fig:xfrac_tau} relies heavily on the assumed value of $\tau_{\rm
  AGN}$, in this section we further explore the X-ray AGN timescale and
the reliability of the estimates we have used.  As a point of reference,
we begin by deriving the value of $\tau_{\rm AGN}$ that would be
necessary to {\em force} the observed quenching and triggering rates to
be equal.  We set $\dot{Q}(M_*) = \dot{\chi}(M_*)$ and solve for
$\tau_{\rm AGN}$ at each stellar mass bin where estimates of the two
rates are available (essentially we divide the full AGN host fraction in
Figure \ref{fig:mfn_xfrac_Lx} by the quenching rates derived in Figure
\ref{fig:edot_frac}).  In principle $\tau_{\rm AGN}$ may be related to
the host stellar mass, but we will ignore this and average over the
values of $\tau_{\rm AGN}$ derived for each mass bin to roughly estimate
the range or ``probability distribution'' of timescales needed to
perfectly match the quenching and AGN trigger rates observed in our two
redshift intervals.  These are plotted as the solid and dashed red lines
in Figure \ref{fig:tau} and illustrate where theoretical estimates of
$\tau_{\rm AGN}$ would fall if it were true that $\dot{Q}(M_*) =
\dot{\chi}(M_*)$.

We can now compare the estimates of $\tau_{\rm AGN}$ we have used and
discuss their uncertainties.  The most appropriate predictions come
from detailed simulations analyzed by Hopkins and collaborators.
\citet{hopkins05a} conduct five hydrodynamical simulations of gas rich
mergers of disk galaxies that host super massive black holes (SMBHs).
During the simulations, gas becomes funneled to the center of the
system, fueling the growth of the newly merged SMBH.  The authors use a
prescription in which some fraction ($\epsilon_r = 0.1$) of this
accreted material is radiated in a quasar phase.  They assume 5\% of
this energy couples to the surrounding gas, helping to regulate the
flow.  The X-ray luminosity (and column density) as a function of time
is calculated by assuming a quasar continuum SED and ray-tracing many
lines of sight through the gas and dust in the simulated galaxy.  Their
Figure 2 presents a relation between the observed AGN lifetime
($\tau_{\rm AGN}$) and its X-ray luminosity.  By applying this relation
to the X-ray luminosities observed in our sample we derive a rough
distribution of predicted AGN timescales which we plot as the dotted
line in Figure \ref{fig:tau}.  Note that the predicted $\tau_{\rm AGN}$
from the \citet{hopkins05a} models increases rapidly for AGNs with $L_X
\lesssim 5 \times 10^{42}$ ergs s$^{-1}$, accounting for the tail
towards longer lifetimes shown in the figure and yielding the estimate
of $\tau_{\rm AGN} \approx 0.6$ Gyr used in Figure \ref{fig:xfrac_tau}.

The work of \citet{granato04} provides another independent comparison, also
based on numerical simulations that encode the effects of star formation,
cooling, supernovae feedback, and AGN feedback set in the context of dark matter
halos.  The systems analyzed in \citet{granato04} are non-interacting spheroidal
galaxies as opposed to merging disks, however, and while the simulations lack detailed modelling of the AGN
X-ray emission, some rough constraints can be obtained for the predicted
values of $\tau_{\rm AGN}$.  Their Figure 3 shows the black hole accretion rate as a
function of cosmic time in their simulations.  Since, as we show below, our AGNs
are likely accreting at significantly sub-Eddington rates, a rough estimate of
$\tau_{\rm AGN}$ from \citet{granato04} can be gained by measuring the time
between peak black hole accretion in their simulations and the point at which the accretion
drops below a factor 0.01--0.001 of the maximum rate.  This yields AGN
timescales between $\sim$1 and $\sim$2 Gyr as shown in Figure \ref{fig:tau} by
the grey shaded region.  For an average value from \citet{granato04} we
take $\tau_{\rm AGN} \approx 1.8$ Gyr.

Finally, much work has been invested in utilizing various observations
to constrain the lifetime of bright quasar activity, typically resulting
in values of 10$^7$--10$^8$ yr \citep[see the review by][]{martini04}.
However, historically the focus has rested on the brightest quasar phase
during which black hole growth is thought to be most rapid.  This phase
does not correspond to the lower accretion rates of the AGNs detected in
our sample.  \citet{marconi04}, however, provide a suitable estimate
based on  matching the local black hole density to
the AGN luminosity function.  For low efficiency sources, the predicted
range is $\tau_{\rm AGN} \approx 0.5--1$ Gyr, weighted more towards 1 Gyr for
the lowest luminosity AGNs and providing a rough average value of $\tau_{\rm
  AGN} = 0.9$ Gyr.

It is clear from Figure \ref{fig:tau} that large uncertainties exist.  Still,
from the detailed analysis in \citet{hopkins05a} to the more
approximate estimates from \citet{marconi04} and \citet{granato04},
this plot demonstrates that the predicted range and
uncertainty in the timescale for X-ray AGN activity using a variety of
methods is at least compatible with a scenario in which AGN triggering is linked
to quenching.  This helps to validate the agreement seen in Figure
\ref{fig:xfrac_tau}.  Clearly further progress in confirming this link would
strongly benefit from additional detailed predictions of the X-ray
properties and lifetimes of AGNs.

\subsection{Is AGN Feedback Responsible For Quenching?}\label{causality}

We have argued that the similarity in the rate of AGN triggering
compared to the rate of star formation quenching suggests that the two
phenomena are linked.  But what is the nature of this link?
Specifically, we would like to know if AGNs, perhaps through feedback
mechanisms, are directly responsible for the quenching of star
formation.  In this section we will begin to probe this question by
investigating the individual properties of our X-ray selected AGNs and how they
correlate with their host galaxies.

Our strategy will be to study the SMBH accretion rates---as
parametrized by the Eddington ratio---in our sample.  We acknowledge that these
estimates are somewhat crude and subject to systematics, but argue that they
nonetheless provide important insight on how AGNs are related to the properties
of their host galaxies.  The Eddington ratio compares the bolometric luminosity,
$L_{\rm bol}$, of the AGN to the Eddington Luminosity which is simply related to
the SMBH mass, $L_{\rm Edd} = 1.25 \times 10^{38} (M_{\rm BH}/M_{\odot})$ ergs
s$^{-1}$.  To determine $M_{\rm BH}$ we use the relation between $K$-band bulge
luminosity and black hole mass calculated by \citet{graham07} and based on
previous work \citep{marconi03, mclure04}.  Note that \citet{woo06} find
evidence that the $z \sim 0.3$ relation is offset relative to that at $z=0$ by a
+0.6 dex increase in $M_{\rm BH}$, although we do not apply this correction here.
As most X-ray selected AGN hosts at the redshifts of our sample have early-type
morphologies \citep{grogin05, pierce07}, we assume that the bulge-to-total ratio
is 1.0, but in what follows we will demonstrate the effect of lower ratios on
our results.  Finally, we calculate $L_{\rm bol}$ by assuming a hard X-ray
bolometric correction of 35 \citep{elvis94}.  \citet{barger05} argue that for
obscured (narrow line) AGNs, the correction should be 85 although
\citet{pozzi07}, using a sample of type-2 AGNs, find a wide range of bolometric
corrections with an average of $\sim$25.  Clearly the bolometric corrections are
uncertain within a factor of $\sim$2--3.

\begin{figure}
\plotone{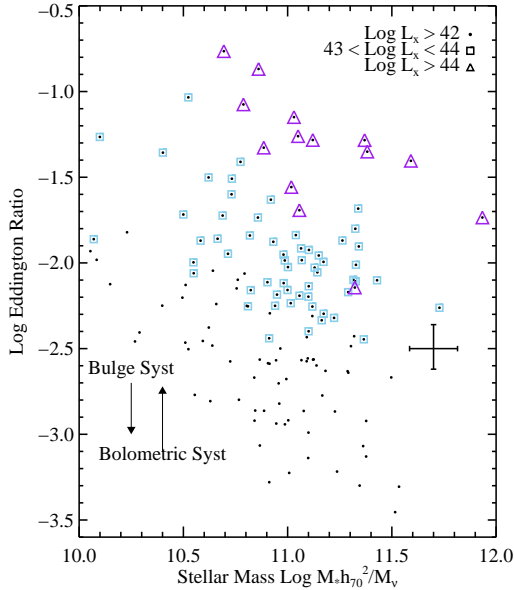}
\caption{Eddington ratios in the AEGIS spec-$z$ sample plotted against the host
  galaxy stellar mass.  Different symbols indicate different ranges of
  AGN X-ray luminosity as indicated.  The isolated error bar illustrates
  the typical uncertainty arising from observational scatter.  The
  labelled arrows suggest the magnitude and sense of systematic
  uncertainties.  The ``Bulge'' systematic shows what would happen if
  the bulge-to-total ratios were lowered by a factor of 2, or
  equivalently if the applied value of $M_{\rm BH}/M_{\rm bulge}$ was
  increased by the same amount.  The ``Bolometric'' systematic shows the
  effect of increasing the bolometric correction from 35 to 85.  Note
  that the axes are not fully independent because the Eddington ratio is
  proportional to $M_*^{-1}$, leading to the slight downward trend
  observed. \label{fig:edd_vs_mass}}
\end{figure}

Using the methods above and with the stated caveats in mind we plot the
estimated Eddington ratio as a function of stellar mass in Figure
\ref{fig:edd_vs_mass}.  The X-ray luminosity range of the data is
indicated, showing that more luminous AGNs tend to have more efficient
accretion.  The isolated error bar indicates the typical uncertainties
from observational scatter, while the arrows show the effects of
systematic errors in estimating the bulge luminosity and bolometric
correction.  The ``Bulge'' systematic shows what would happen if the
bulge-to-total ratios were lowered by a factor of 2, or equivalently if
the applied value of $M_{\rm BH}/M_{\rm bulge}$ was increased by the same
amount.  The ``Bolometric'' systematic shows the effect of increasing
the bolometric correction from 35 to 85.  It is important to note that
our calculation of the Eddington ratio ensures that it is proportional
to $M_*^{-1}$ so that downward mass-dependent trends in Figure
\ref{fig:edd_vs_mass} are expected given our methodology.

\begin{figure}
\plotone{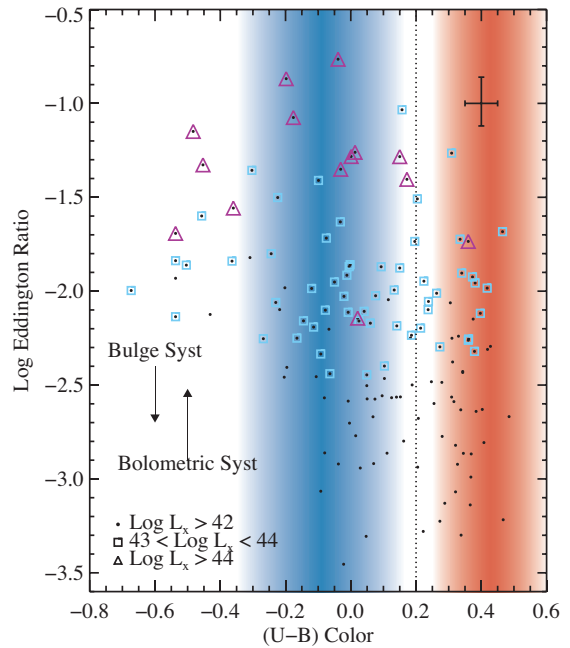}
\caption{Eddington ratios as in Figure \ref{fig:edd_vs_mass} for host
  galaxies of varying restframe $(U-B)$ color.  A typical value for the
  division between red and blue galaxies is shown with the vertical
  dotted line and the color distributions of all galaxies is
  indicated by the background blue and red shading. \label{fig:edd_vs_col}}
\end{figure}

What is perhaps most revealing about the figure, however, is the large
range (more than 2 orders of magnitude) in Eddington ratios that is
apparent for host galaxies of all masses.  Figure \ref{fig:edd_vs_col}
shows a similar diagram where it is now possible to compare the
Eddington ratios versus host galaxy restframe ($U-B$) color.  The red
and blue distributions of all galaxies are indicated by the
background shading.  Note the existence of very blue hosts (beyond the
range of normal colors) dominated by bright X-ray systems.  As observed
in \citet{nandra07}, it is likely that the AGN contaminates the host
color of these galaxies.

Considering the hosts in the blue cloud and red sequence, there appears
to be little difference in the large spread of accretion rates.  This is
especially the case when one ignores the X-ray brightest systems (marked
as triangles), whose true host colors may be redder than observed.  We
note that among all galaxies at $z \approx 0$ the fraction of dusty but
still star-forming systems on the red sequence is $\sim$7\% and is not
likely to be significantly higher at $z \sim 1$ \citep{yan06}.  It is
therefore likely that most red AGN hosts in Figure \ref{fig:edd_vs_col}
are truly quenched systems.

While the uncertainties in Figures \ref{fig:edd_vs_mass} and
\ref{fig:edd_vs_col} are large, the intrinsic scatter appears to be more
substantial and suggests little or no additional trends with stellar
mass or color beyond those expected from the methodology.  These
observations may therefore have important implications for the question
of whether AGN feedback is responsible for quenching.  

Some feedback models imagine that the AGN has an explosive episode that
drives gas out of the galaxy halo.  In the model of Hopkins and
collaborators discussed above, for example, AGN activity is triggered by
major mergers.  The buried AGN is virtually undetectable until a violent
quasar phase in which tremendous energy is expelled, sufficient to heat
or dispel most of the remaining gas in the galaxy, thus quenching
further star formation.  As described in \citet{hopkins07}, in this
model one would expect that AGN activity evolves with time as AGN
feedback during the quasar phase impacts the surrounding material,
heating and driving it from the newly merged system.  The X-ray
observations studied here would then correspond to the post-quasar phase
tracing AGNs as they decay to low luminosities; beforehand the
obscuration is predicted to be $\gtrsim$10$^{24}$ cm$^{-2}$, enough to
absorb hard X-rays.  The short-lived quasar phase that immediately follows would be
too bright to enable studies of host properties
\citep[see][]{hopkins07}.

Figures \ref{fig:edd_vs_mass} and \ref{fig:edd_vs_col} suggest some
potential problems with simple interpretations of explosive models of
this sort.  One might expect the most massive and reddest hosts to have
been quenched earliest and therefore to harbor AGNs in the latest stages
of decay.  Little fuel should remain in such systems long after the
quasar phase, and the luminosities and Eddington ratios should be low.
This picture appears consistent with observations of local AGN hosts
identified through optical emission line diagnostics.  Stronger emission
line AGNs are found in younger stellar populations with higher specific
star formation rates, while weaker AGNs favor hosts that have apparently
been quenched \citep{kauffmann03b, salim07}.  Figures
\ref{fig:edd_vs_mass} and \ref{fig:edd_vs_col} show that X-ray selected
AGNs may present a different picture.  Here, the reddest and most
massive hosts harbor AGNs that cover nearly the full range of X-ray
luminosity and accretion rates.  Indeed, Figure \ref{fig:mfn_x}
demonstrates that AGNs are hosted by galaxies with masses covering the
full range probed.

One perspective on this question is suggested by the work of
\citet{ciotti07} who demonstrate that X-ray luminous AGN activity as
well as starbursts can be effectively fueled by stellar mass loss from
evolved stars in old stellar populations.  While it is not clear what
timescales would be involved, in the absence of other fueling
mechanisms, this process requires an old stellar population to function,
generating an obvious link between quenching and the appearance of AGNs.
The work of \citet{ciotti07} serves to demonstrate that AGNs may be
fueled by a variety of mechanisms.  Even if a high-accretion quasar
phase was initially responsible for disrupting the internal gas supply,
the AGN may later be ``refueled'' by such mechanisms, including the
inflow of gas lost from evolved stars.  

Finally, in addition to refueling, an alternative explanation for the
range of AGN properties observed in the most massive and reddest hosts
could come from the notion that AGN feedback is {\em not} responsible
for quenching star formation, but is triggered by the same process that
is.  A nuclear starburst fueled by the same inflowing gas that ignites
the AGN is a promising example, capable of providing the feedback energy
necessary (in the form of stellar winds) to heat and expel the
surrounding gas supply and help regulate correlations between bulge and
SMBH properties \citep[see][]{ferrarese05}.  Previous work has revealed
evidence for a connection between starbursts and AGN activity
\citep[e.g.,][]{yan06, goto06, yang06, wild07, georgakakis07a} which has
also been explored in models \citep[e.g.,][]{somerville01, hopkins07a}.

A mixed AGN/starburst scenario is supported by the observations
presented here.  AGN activity could be initially triggered during or
towards the end of starburst quenching, leading to a greater diversity
in the phases of X-ray detected AGN activity among galaxy hosts.  As
suggested by \citet{croton06}, low luminosity AGNs---undetected in the
current sample---may be ubiquitous in quenched systems
\citep[e.g.,][]{salim07}, providing the necessary feedback that prevents
further star formation.  As part of this feedback cycle, these systems
could periodically enter active phases that could be detected through
X-ray emission.

\section{ Summary}\label{summary}

We have used the combination of the DEEP2 Galaxy Redshift Survey, Palomar near-IR
imaging, and {\it Chandra} X-ray observations to study the properties of
galaxies that host X-ray selected AGNs.  We summarize our findings below.

\begin{itemize}

\item The AGN host stellar mass function over the redshift range $0.4 <
  z < 1.4$ is roughly flat as a function of $M_*$.  The abundance of
  AGNs appears higher at $z \sim 1$ by a factor of $\sim$2.  Coupled
  with the decrease in number density of all galaxies at high redshift,
  the AGN fraction increases at early times, especially among the most
  massive galaxies.

\item The MF of host galaxies for an X-ray luminous subset of our sample with AGN
  luminosities of $L_{2-10} > 10^{43}$ ergs s$^{-1}$ indicates a lower abundance
  but a very similar mass dependence as the full sample with $L_{2-10} >
  10^{42}$ ergs s$^{-1}$.  We see some evidence, however, that brighter AGNs,
  whose abundance increases with redshift, are hosted by more massive galaxies.

\item Using the full DEEP2/Palomar sample we estimate the star formation
  quenching rate, defined as the number of galaxies that move to the red sequence
  per Gyr.  Our estimates suggest that massive galaxies not only populated the
  red sequence at earlier epochs but did so at a faster rate than less massive galaxies.

\item We show that the quenching rate agrees with the rate at which AGN activity is
  triggered in galaxies if the lifetime over which AGNs would be detected
  through X-ray emission is 0.5--1 Gyr, similar to estimates from detailed predictions
  based on numerical simulations and black hole demographics.  The agreement between the mass-dependent
  quenching and AGN triggering rates is evidence of a physical link between these two
  phenomena.

\item We test the causality of this link by comparing black hole
  accretion rates to the stellar mass and color of associated host
  galaxies.  The most massive and red hosts---which presumably have
  quenched at the earliest times---harbor X-ray selected AGNs as active
  as those found in blue hosts.  This suggests a more complicated
  relationship between AGNs and star formation. It is possible that
  X-ray selected AGNs are associated with but do not directly {\em
    cause} star formation quenching and, furthermore, may be subject to
  refueling after quenching occurs.

\end{itemize}

\acknowledgments

We would like to thank Ray Carlberg for useful discussions on this
project.  KB would like to acknowledge support from Olivier Le F\`{e}vre
and the Observatoire Astronomique de Marseille Provence where he was a
visiting researcher. The Palomar Survey was supported by NSF grant
AST-0307859 and NASA STScI grant HST-AR-09920.01-A.  Support from
National Science Foundation grants 00-71198 to UCSC and AST~00-71048 to
UCB is also gratefully acknowledged.  Financial support has also been
provided through PPARC and the Marie Curie Fellowship grant
MEIF-CT-2005-025108 (AG), the Leverhulme trust (KN), the Hubble
Fellowship grants HF-01165.01-A (JAN) and HF-01182.01-A (ALC) and the
STFC (EL).  We wish to recognize and acknowledge the highly significant
cultural role and reverence that the summit of Mauna Kea has always had
within the indigenous Hawaiian community. It is a privilege to be given
the opportunity to conduct observations from this mountain.



\begin{thebibliography}{77}
\expandafter\ifx\csname natexlab\endcsname\relax\def\natexlab#1{#1}\fi

\bibitem[{{Barger} {et~al.}(2005){Barger}, {Cowie}, {Mushotzky}, {Yang},
  {Wang}, {Steffen}, \& {Capak}}]{barger05}
{Barger}, A.~J., {Cowie}, L.~L., {Mushotzky}, R.~F., {Yang}, Y., {Wang}, W.-H.,
  {Steffen}, A.~T., \& {Capak}, P. 2005, \aj, 129, 578

\bibitem[{{Bauer} {et~al.}(2004){Bauer}, {Alexander}, {Brandt}, {Schneider},
  {Treister}, {Hornschemeier}, \& {Garmire}}]{bauer04}
{Bauer}, F.~E., {Alexander}, D.~M., {Brandt}, W.~N., {Schneider}, D.~P.,
  {Treister}, E., {Hornschemeier}, A.~E., \& {Garmire}, G.~P. 2004, \aj, 128,
  2048

\bibitem[{{Bell} {et~al.}(2004){Bell}, {Wolf}, {Meisenheimer}, {Rix}, {Borch},
  {Dye}, {Kleinheinrich}, {Wisotzki}, \& {McIntosh}}]{bell04}
{Bell}, E.~F., {Wolf}, C., {Meisenheimer}, K., {Rix}, H.-W., {Borch}, A.,
  {Dye}, S., {Kleinheinrich}, M., {Wisotzki}, L., \& {McIntosh}, D.~H. 2004,
  \apj, 608, 752

\bibitem[{{Ben{\'{\i}}tez}(2000)}]{benitez00}
{Ben{\'{\i}}tez}, N. 2000, \apj, 536, 571

\bibitem[{{Bertin} \& {Arnouts}(1996)}]{bertin96}
{Bertin}, E. \& {Arnouts}, S. 1996, \aaps, 117, 393

\bibitem[{{Borch} {et~al.}(2006)}]{borch06}
{Borch}, A. {et~al.} 2006, \aap, 453, 869

\bibitem[{{Bower} {et~al.}(2006)}]{bower06}
{Bower}, R.~G. {et~al.} 2006, \mnras, 370, 645

\bibitem[{{Bruzual} \& {Charlot}(2003)}]{bruzual03}
{Bruzual}, G. \& {Charlot}, S. 2003, \mnras, 344, 1000

\bibitem[{{Bundy} {et~al.}(2005){Bundy}, {Ellis}, \& {Conselice}}]{bundy05}
{Bundy}, K., {Ellis}, R.~S., \& {Conselice}, C.~J. 2005, \apj, 625, 621

\bibitem[{{Bundy} {et~al.}(2006)}]{bundy06}
{Bundy}, K. {et~al.} 2006, \apj, 651, 120

\bibitem[{{Canalizo} \& {Stockton}(2001)}]{canalizo01}
{Canalizo}, G. \& {Stockton}, A. 2001, \apj, 555, 719

\bibitem[{{Cattaneo} {et~al.}(2006){Cattaneo}, {Dekel}, {Devriendt},
  {Guiderdoni}, \& {Blaizot}}]{cattaneo06}
{Cattaneo}, A., {Dekel}, A., {Devriendt}, J., {Guiderdoni}, B., \& {Blaizot},
  J. 2006, \mnras, 370, 1651

\bibitem[{{Chabrier}(2003)}]{chabrier03}
{Chabrier}, G. 2003, \pasp, 115, 763

\bibitem[{{Cimatti} {et~al.}(2006){Cimatti}, {Daddi}, \& {Renzini}}]{cimatti06}
{Cimatti}, A., {Daddi}, E., \& {Renzini}, A. 2006, \aap, 453, L29

\bibitem[{{Ciotti} \& {Ostriker}(2007)}]{ciotti07}
{Ciotti}, L. \& {Ostriker}, J.~P. 2007, \apj, 665, 1038

\bibitem[{{Coil} {et~al.}(2004){Coil}, {Newman}, {Kaiser}, {Davis}, {Ma},
  {Kocevski}, \& {Koo}}]{coil04}
{Coil}, A.~L., {Newman}, J.~A., {Kaiser}, N., {Davis}, M., {Ma}, C.-P.,
  {Kocevski}, D.~D., \& {Koo}, D.~C. 2004, \apj, 617, 765

\bibitem[{{Collister} \& {Lahav}(2004)}]{collister04}
{Collister}, A.~A. \& {Lahav}, O. 2004, \pasp, 116, 345

\bibitem[{{Conselice} {et~al.}(2007)}]{conselice07}
{Conselice}, C.~J. {et~al.} 2007, preprint (arXiv0708.1040), 708

\bibitem[{{Cooper} {et~al.}(2007)}]{cooper07}
{Cooper}, M.~C. {et~al.} 2007, preprint (arXiv0706.4089), 706

\bibitem[{{Cowie} {et~al.}(1996){Cowie}, {Songaila}, {Hu}, \&
  {Cohen}}]{cowie96}
{Cowie}, L.~L., {Songaila}, A., {Hu}, E.~M., \& {Cohen}, J.~G. 1996, \aj, 112,
  839

\bibitem[{{Croton} {et~al.}(2006)}]{croton06}
{Croton}, D.~J. {et~al.} 2006, \mnras, 365, 11

\bibitem[{{Cuillandre} {et~al.}(2001){Cuillandre}, {Luppino}, {Starr}, \&
  {Isani}}]{cuillandre01}
{Cuillandre}, J.-C., {Luppino}, G., {Starr}, B., \& {Isani}, S. 2001, in
  SF2A-2001: Semaine de l'Astrophysique Francaise, 605--+

\bibitem[{{Davis} {et~al.}(2005){Davis}, {Gerke}, {Newman}, \& {the Deep2
  Team}}]{davis05}
{Davis}, M., {Gerke}, B.~F., {Newman}, J.~A., \& {the Deep2 Team}. 2005, in ASP
  Conf. Ser. 339: Observing Dark Energy, ed. S.~C. {Wolff} \& T.~R. {Lauer},
  128--+

\bibitem[{{Davis} {et~al.}(2003)}]{davis03}
{Davis}, M. {et~al.} 2003, in Discoveries and Research Prospects from 6- to
  10-Meter-Class Telescopes II. Edited by Guhathakurta, Puragra. Proceedings of
  the SPIE, Volume 4834, pp. 161-172 (2003)., 161--172

\bibitem[{{Davis} {et~al.}(2007)}]{davis07}
{Davis}, M. {et~al.} 2007, \apjl, 660, L1

\bibitem[{{de Lucia} \& {Blaizot}(2007)}]{de-lucia07}
{de Lucia}, G. \& {Blaizot}, J. 2007, \mnras, 375, 2

\bibitem[{{Dunlop} {et~al.}(2003){Dunlop}, {McLure}, {Kukula}, {Baum}, {O'Dea},
  \& {Hughes}}]{dunlop03}
{Dunlop}, J.~S., {McLure}, R.~J., {Kukula}, M.~J., {Baum}, S.~A., {O'Dea},
  C.~P., \& {Hughes}, D.~H. 2003, \mnras, 340, 1095

\bibitem[{{Elvis} {et~al.}(1994)}]{elvis94}
{Elvis}, M. {et~al.} 1994, \apjs, 95, 1

\bibitem[{{Faber} {et~al.}(2007)}]{faber07}
{Faber}, S.~M. {et~al.} 2007, \apj, 665, 265

\bibitem[{{Ferrarese} \& {Ford}(2005)}]{ferrarese05}
{Ferrarese}, L. \& {Ford}, H. 2005, Space Science Reviews, 116, 523

\bibitem[{{Feulner} {et~al.}(2005){Feulner}, {Gabasch}, {Salvato}, {Drory},
  {Hopp}, \& {Bender}}]{feulner05a}
{Feulner}, G., {Gabasch}, A., {Salvato}, M., {Drory}, N., {Hopp}, U., \&
  {Bender}, R. 2005, \apjl, 633, L9

\bibitem[{{Georgakakis} {et~al.}(2006)}]{georgakakis06}
{Georgakakis}, A. {et~al.} 2006, \mnras, 371, 221

\bibitem[{{Georgakakis} {et~al.}(2007{\natexlab{a}})}]{georgakakis07a}
---. 2007{\natexlab{a}}, in prep., 660, L15

\bibitem[{{Georgakakis} {et~al.}(2007{\natexlab{b}})}]{georgakakis07}
---. 2007{\natexlab{b}}, \apjl, 660, L15

\bibitem[{{Gerke} {et~al.}(2007)}]{gerke07}
{Gerke}, B.~F. {et~al.} 2007, \mnras, 376, 1425

\bibitem[{{Gilli} {et~al.}(2007){Gilli}, {Comastri}, \& {Hasinger}}]{gilli07}
{Gilli}, R., {Comastri}, A., \& {Hasinger}, G. 2007, \aap, 463, 79

\bibitem[{{Goto}(2006)}]{goto06}
{Goto}, T. 2006, \mnras, 369, 1765

\bibitem[{{Graham}(2007)}]{graham07}
{Graham}, A.~W. 2007, \mnras, 543

\bibitem[{{Granato} {et~al.}(2004){Granato}, {De Zotti}, {Silva}, {Bressan}, \&
  {Danese}}]{granato04}
{Granato}, G.~L., {De Zotti}, G., {Silva}, L., {Bressan}, A., \& {Danese}, L.
  2004, \apj, 600, 580

\bibitem[{{Grogin} {et~al.}(2005)}]{grogin05}
{Grogin}, N.~A. {et~al.} 2005, \apjl, 627, L97

\bibitem[{{Guainazzi} {et~al.}(2005){Guainazzi}, {Matt}, \&
  {Perola}}]{guainazzi05}
{Guainazzi}, M., {Matt}, G., \& {Perola}, G.~C. 2005, \aap, 444, 119

\bibitem[{{Hasinger} {et~al.}(2005){Hasinger}, {Miyaji}, \&
  {Schmidt}}]{hasinger05}
{Hasinger}, G., {Miyaji}, T., \& {Schmidt}, M. 2005, \aap, 441, 417

\bibitem[{{Hopkins} {et~al.}(2007{\natexlab{a}}){Hopkins}, {Cox}, {Keres}, \&
  {Hernquist}}]{hopkins07a}
{Hopkins}, P.~F., {Cox}, T.~J., {Keres}, D., \& {Hernquist}, L.
  2007{\natexlab{a}}, preprint (arXiv0706.1246), 706

\bibitem[{{Hopkins} {et~al.}(2007{\natexlab{b}}){Hopkins}, {Hernquist}, {Cox},
  \& {Keres}}]{hopkins07}
{Hopkins}, P.~F., {Hernquist}, L., {Cox}, T.~J., \& {Keres}, D.
  2007{\natexlab{b}}, preprint (arXiv0706.1243), 706

\bibitem[{{Hopkins} {et~al.}(2005{\natexlab{a}}){Hopkins}, {Hernquist},
  {Martini}, {Cox}, {Robertson}, {Di Matteo}, \& {Springel}}]{hopkins05b}
{Hopkins}, P.~F., {Hernquist}, L., {Martini}, P., {Cox}, T.~J., {Robertson},
  B., {Di Matteo}, T., \& {Springel}, V. 2005{\natexlab{a}}, \apjl, 625, L71

\bibitem[{{Hopkins} {et~al.}(2005{\natexlab{b}})}]{hopkins05a}
{Hopkins}, P.~F. {et~al.} 2005{\natexlab{b}}, \apj, 630, 705

\bibitem[{{Hutchings} {et~al.}(2006){Hutchings}, {Cherniawsky}, {Cutri}, \&
  {Nelson}}]{hutchings06}
{Hutchings}, J.~B., {Cherniawsky}, A., {Cutri}, R.~M., \& {Nelson}, B.~O. 2006,
  \aj, 131, 680

\bibitem[{{Ilbert} {et~al.}(2006)}]{ilbert06}
{Ilbert}, O. {et~al.} 2006, \aap, 457, 841

\bibitem[{{Juneau} {et~al.}(2005)}]{juneau05}
{Juneau}, S. {et~al.} 2005, \apjl, 619, L135

\bibitem[{{Kannappan} \& {Gawiser}(2007)}]{kannappan07}
{Kannappan}, S.~J. \& {Gawiser}, E. 2007, \apjl, 657, L5

\bibitem[{{Kauffmann} {et~al.}(2003)}]{kauffmann03b}
{Kauffmann}, G. {et~al.} 2003, \mnras, 346, 1055

\bibitem[{{Maraston} {et~al.}(2006)}]{maraston06}
{Maraston}, C. {et~al.} 2006, \apj, 652, 85

\bibitem[{{Marconi} \& {Hunt}(2003)}]{marconi03}
{Marconi}, A. \& {Hunt}, L.~K. 2003, \apjl, 589, L21

\bibitem[{{Marconi} {et~al.}(2004){Marconi}, {Risaliti}, {Gilli}, {Hunt},
  {Maiolino}, \& {Salvati}}]{marconi04}
{Marconi}, A., {Risaliti}, G., {Gilli}, R., {Hunt}, L.~K., {Maiolino}, R., \&
  {Salvati}, M. 2004, \mnras, 351, 169

\bibitem[{{Martin} {et~al.}(2007)}]{martin07}
{Martin}, D.~C. {et~al.} 2007, preprint (astro-ph/0703281)

\bibitem[{{Martini}(2004)}]{martini04}
{Martini}, P. 2004, in Coevolution of Black Holes and Galaxies, ed. L.~C. {Ho},
  169--+

\bibitem[{{McLure} \& {Dunlop}(2004)}]{mclure04}
{McLure}, R.~J. \& {Dunlop}, J.~S. 2004, \mnras, 352, 1390

\bibitem[{{Mushotzky}(2004)}]{mushotzky04}
{Mushotzky}, R. 2004, in Astrophysics and Space Science Library, Vol. 308,
  Astrophysics and Space Science Library, ed. A.~J. {Barger}, 53--+

\bibitem[{{Nandra} \& {Pounds}(1994)}]{nandra94}
{Nandra}, K. \& {Pounds}, K.~A. 1994, \mnras, 268, 405

\bibitem[{{Nandra} {et~al.}(2005)}]{nandra05}
{Nandra}, K. {et~al.} 2005, \mnras, 356, 568

\bibitem[{{Nandra} {et~al.}(2007)}]{nandra07}
---. 2007, \apjl, 660, L11

\bibitem[{{Pierce} {et~al.}(2007)}]{pierce07}
{Pierce}, C.~M. {et~al.} 2007, \apjl, 660, L19

\bibitem[{{Pozzi} {et~al.}(2007)}]{pozzi07}
{Pozzi}, F. {et~al.} 2007, \aap, 468, 603

\bibitem[{{Salim} {et~al.}(2007)}]{salim07}
{Salim}, S. {et~al.} 2007, preprint (0704.3611), 704

\bibitem[{{Scannapieco} {et~al.}(2005){Scannapieco}, {Silk}, \&
  {Bouwens}}]{scannapieco05}
{Scannapieco}, E., {Silk}, J., \& {Bouwens}, R. 2005, \apjl, 635, L13

\bibitem[{{Schmidt}(1968)}]{schmidt68}
{Schmidt}, M. 1968, \apj, 151, 393

\bibitem[{{Silk} \& {Rees}(1998)}]{silk98}
{Silk}, J. \& {Rees}, M.~J. 1998, \aap, 331, L1

\bibitem[{{Somerville} {et~al.}(2001){Somerville}, {Primack}, \&
  {Faber}}]{somerville01}
{Somerville}, R.~S., {Primack}, J.~R., \& {Faber}, S.~M. 2001, \mnras, 320, 504

\bibitem[{{Treister} \& {Urry}(2006)}]{treister06}
{Treister}, E. \& {Urry}, C.~M. 2006, \apjl, 652, L79

\bibitem[{{Treu} {et~al.}(2005){Treu}, {Ellis}, {Liao}, \& {van
  Dokkum}}]{treu05}
{Treu}, T., {Ellis}, R.~S., {Liao}, T.~X., \& {van Dokkum}, P.~G. 2005, \apjl,
  622, L5

\bibitem[{{van Dokkum} {et~al.}(2000){van Dokkum}, {Franx}, {Fabricant},
  {Illingworth}, \& {Kelson}}]{van-dokkum00}
{van Dokkum}, P.~G., {Franx}, M., {Fabricant}, D., {Illingworth}, G.~D., \&
  {Kelson}, D.~D. 2000, \apj, 541, 95

\bibitem[{{Wild} {et~al.}(2007){Wild}, {Kauffmann}, {Heckman}, {Charlot},
  {Lemson}, {Brinchmann}, {Reichard}, \& {Pasquali}}]{wild07}
{Wild}, V., {Kauffmann}, G., {Heckman}, T., {Charlot}, S., {Lemson}, G.,
  {Brinchmann}, J., {Reichard}, T., \& {Pasquali}, A. 2007, preprint
  (arXiv0706.3113), 706

\bibitem[{{Willmer} {et~al.}(2006)}]{willmer06}
{Willmer}, C.~N.~A. {et~al.} 2006, \apj, 647, 853

\bibitem[{{Wilson} {et~al.}(2003)}]{wilson03}
{Wilson}, J.~C. {et~al.} 2003, in Instrument Design and Performance for
  Optical/Infrared Ground-based Telescopes. Edited by Iye, Masanori; Moorwood,
  Alan F. M. Proceedings of the SPIE, Volume 4841, pp. 451-458 (2003).,
  451--458

\bibitem[{{Woo} {et~al.}(2006){Woo}, {Treu}, {Malkan}, \& {Blandford}}]{woo06}
{Woo}, J.-H., {Treu}, T., {Malkan}, M.~A., \& {Blandford}, R.~D. 2006, \apj,
  645, 900

\bibitem[{{Yan} {et~al.}(2006){Yan}, {Newman}, {Faber}, {Konidaris}, {Koo}, \&
  {Davis}}]{yan06}
{Yan}, R., {Newman}, J.~A., {Faber}, S.~M., {Konidaris}, N., {Koo}, D., \&
  {Davis}, M. 2006, \apj, 648, 281

\bibitem[{{Yang} {et~al.}(2006){Yang}, {Tremonti}, {Zabludoff}, \&
  {Zaritsky}}]{yang06}
{Yang}, Y., {Tremonti}, C.~A., {Zabludoff}, A.~I., \& {Zaritsky}, D. 2006,
  \apjl, 646, L33

\end{thebibliography}

\end{document}